\documentclass[
aps,%
12pt,%
final,%
notitlepage,%
oneside,%
onecolumn,%
nobibnotes,%
nofootinbib,%
superscriptaddress,%
showpacs,%
centertags]%
{revtex4}

\usepackage[english]{babel}
\usepackage{epsfig,amssymb}

\def\bb {\begin {eqnarray}}
\def\ee {\end {eqnarray}}

\begin{document}

\title
{Supersymmetric partners for the \\ associated Lam\'e
potentials
}

\author{\firstname{David J.}~\surname{Fern\'andez C.}}
\affiliation{Departamento de F\'{\i}sica, Cinvestav, \\ A.P. 14-740,
07000 M\'exico D.F., Mexico}

\author{\firstname{Asish}~\surname{Ganguly}}
\affiliation{City College, University of Calcutta, 13 Surya Sen
Street, Kolkata-700012, India}

\begin{abstract}
\noindent The general solution of the stationary Schr\"odinger
equation for the associated Lam\'e potentials with an arbitrary real
energy is found. The supersymmetric partners are generated by
employing seeds solutions for factorization energies inside the
gaps.
\end{abstract}


\maketitle

\section{Introduction}

In quantum mechanics the exactly solvable models (ESM) are
essential: since the complete physical information is encoded in few
analytic expressions, they are ideal to test the convergence of
numerical methods. Moreover, the ESM constitute the starting point
for applying perturbation techniques. There are some other models,
intermediate between the exactly solvable ones and those which can
be solved just numerically, which are known nowadays as
quasi-exactly solvable (QES). For them there exist analytical
expressions encoding just partial physical information, e.g., for
only a part of the Hamiltonian spectrum \cite{us94}. Along the years
several systems have been identified as belonging to the QES class,
in particular, there was a dominant conviction that the associated
Lam\'e potentials were QES \cite{rkp05}. However, very recently it
was shown that the associated Lam\'e potentials for $(m,\ell)$
integers are exactly solvable, in the sense that the stationary
Schr\"odinger equation admits analytic solutions for any value of
the energy parameter \cite{fg05,fg07}. The initial motivation to
look for this result was the need to implement supersymmetric
quantum mechanics for generating new exactly solvable models
(periodic and asymptotically periodic)
\cite{df98,ks99,fnn00,fmrs02a,fmrs02b,sgnn03}. In order to implement
non-singular transformations of general type, the explicit
expressions for the Schr\"odinger solutions in the gaps as well as
the band edge eigenfunctions were required \cite{fmrs02a,fmrs02b}.

In the next section we will discuss two techniques to find the
general solution for the associated Lam\'e equation for an arbitrary
value of the energy parameter: first an ansatz based procedure and
then a systematic technique related to the well known Frobenius
method. In section 3 we will apply the supersymmetry transformations
for generating new periodic and asymptotically periodic potentials
which are almost isospectral to the initial associated Lam\'e
potential. In section 4 our general results will be illustrated
through the particular case characterized by $(m,\ell) = (3,2)$. Our
conclusions will be finally given at section 5.

\section{General solutions of the associated Lam\'e equation}

We want to find the general solution of the Schr\"odinger equation
for the associated Lam\'e potentials with an arbitrary value of the
energy parameter $E$:
\begin{eqnarray}
&& H \psi(x) = \left[-\partial^2_x + V(x)\right]\psi(x)=E\psi(x), \\
&& V(x) = m(m+1)k^2{\rm sn}^2x+\ell(\ell+1)k^2\frac{{\rm
cn}^2x}{{\rm dn}^2x}, \label{alp}
\end{eqnarray}
where $m,\ell\in{\mathbb N}, \ m\geq \ell$, and ${\rm sn}x \equiv
{\rm sn}(x,k)$, ${\rm cn}x\equiv {\rm cn}(x,k)$, ${\rm dn}x\equiv
{\rm dn}(x,k)$ are the Jacobi elliptic functions of real periods
$4K, \ 4K$ and $2K$ respectively \cite{fg05,fg07}. This is the
associated Lam\'e equation, which can be also expressed in
Weierstrass form:
\begin{eqnarray}
&& \label{aw} -\frac{d^2\psi}{dz^2}+ \left[m(m+1)\wp(z)+
\frac{\ell(\ell+1)\bar{e}_2\bar{e}_3}{\wp(z)-e_1}\right]\psi=\widetilde{E}\psi,
\label{wfale}
\end{eqnarray}
through the changes
\begin{eqnarray}
\nonumber && z=\frac{x-iK'}{\sqrt{\bar{e}_{3}}}, \quad
\bar{e}_i=e_1-e_i, \quad i=2,3, \quad
\widetilde{E}=e_3m(m+1)+[E-\ell(\ell+1)]\bar{e}_3,
\end{eqnarray}
$\wp(z)\equiv \wp(z,\omega,\omega')$ being the Weierstrass elliptic
function of half-periods $\omega = K/\sqrt{\bar{e}_3}$, $\omega' = i
K'/\sqrt{\bar{e}_3}$.

In order to solve (\ref{wfale}), let us denote by $\psi^\pm(z)$ two
linearly independent solutions. Thus, their product
$\Psi(z)=\psi^+\psi^-$ will satisfy the following third-order
equation:
\begin{eqnarray}
&& \hspace{-1cm} \frac{d^3\Psi}{dz^3}-4\left[m(m+1)\wp(z)+\frac{\ell(\ell+1)
\bar{e}_2\bar{e}_3}{\wp(z)-e_1}- \widetilde{E}\right]
\frac{d\Psi}{dz} \nonumber \\
\label{prod1} &&
-2\left\{m(m+1)-\frac{\ell(\ell+1)\bar{e}_2\bar{e}_3}{[
\wp(z)-e_1]^2}\right\}\wp'(z)\Psi = 0.
\end{eqnarray}

In the first place let us consider the following ansatz for $\Psi$:
\begin{eqnarray}
&& \Psi(z)=[\wp(z)-e_{1}]+A_{1} + \frac{A_{2}}{\wp(z)-e_{1}} .
\end{eqnarray}
By plugging this in equation (\ref{prod1}), the parameters $A_1, \
A_2, \ \ell, m$ can be fitted \cite{fg05}:
\begin{eqnarray*}
(a) \quad m=1, \quad \ell=1, & \quad A_{1}=\widetilde{E} + e_{1},
                & \quad A_{2}=\bar{e}_{2}\bar{e}_{3},
\label{sol1,1} \\
(b) \quad m=1, \quad \ell=0, & \quad A_{1}=\widetilde{E} + e_{1},
                    & \quad A_{2} = 0. \label{sol1,2}
\end{eqnarray*}

Once $\Psi(z)$ has been gotten, it is straightforward to find
$\psi^{\pm}(z)$ through the formula:
\begin{eqnarray}
\label{usp} && \psi^{\pm}(z)=\sqrt{\Psi(z)}\exp\left
(\mp\frac{1}{2}\int^z\frac{d\tau}{\Psi(\tau)}\right).
\end{eqnarray}
In the non-trivial case (a) with $(m,\ell)=(1,1)$, in which the associated
Lam\'e equation is not directly the Lam\'e one, the corresponding solution
reads:
\begin{eqnarray*}
\psi_{\pm}(x) & = & \frac{\prod\limits^{2}_{r=1}
\sigma\left(\frac{x-iK'}{\sqrt{\bar{e}_{3}}} \pm a_{r} \right)}{
\sigma\left(\frac{x-iK'}{\sqrt{\bar{e}_{3}}}+ \omega_1\right)
\sigma\left(\frac{x-iK'}{\sqrt{\bar{e}_{3}}}\right)}
\exp\left[\mp\frac{x}{\sqrt{\bar{e}_{3}}} \sum\limits^{2}_{r=0}\zeta
(a_{r})\right] ,
\end{eqnarray*}
where $\omega_1 = \omega$, $\zeta'(z)=-\wp(z), \
[\ln\sigma(z)]'=\zeta(z)$.

On the other hand, for a modified ansatz of kind \cite{fg05}:
\begin{eqnarray*}
\Psi(z)=[\wp(z)-e_{1}]^{2}+B_{1}[\wp(z)-e_{1}]+B_{2} +
\frac{B_{3}}{\wp(z)-e_{1}},  \label{ansz2}
\end{eqnarray*}
it turns out that
\begin{eqnarray*}
(c) \: m = 2, \quad \ell  =  1, && B_{1} =  2e_{1}  +
\frac{\widetilde{E}}3, \quad  B_{2} = \left(\frac{\widetilde{E}}3  -
e_{1}\right) B_{1}, \quad B_{3} =
\frac{\bar{e}_{2}\bar{e}_{3}B_{1}}3,
\label{sol2,1} \\
(d) \: m  =  2, \quad \ell  =  0, && B_{1} =  2e_{1}  +
\frac{\widetilde{E}}3, \quad B_{2} = \left(\frac{\widetilde{E}}3  -
e_{1}\right) B_{1} + \bar{e}_{2}\bar{e}_{3}, \quad B_{3}  =  0
\label{sol2,2}.
\end{eqnarray*}
Once again, in the case (c) with $(m,\ell)=(2,1)$, in which the associated
Lam\'e equation is not the Lam\'e one, the corresponding $\psi$ solutions are:
\begin{eqnarray*}
\psi_{\pm}(x) & = & \frac{\prod\limits^{3}_{r=1}
\sigma\left(\frac{x-iK'}{\sqrt{\bar{e}_{3}}} \pm b_{r} \right)}{
\sigma\left(\frac{x-iK'}{\sqrt{\bar{e}_{3}}}+ \omega_1\right)
\sigma^2\left(\frac{x-iK'}{\sqrt{\bar{e}_{3}}}\right)} \exp
\left[\mp\frac{x}{\sqrt{\bar{e}_{3}}} \sum\limits^{3}_{r=0}\zeta
(b_{r})\right].  \label{gs2,1,1}
\end{eqnarray*}

Although the previous approach allows to find the solutions of equations
(\ref{wfale}-\ref{prod1}) for some integer values of the parameters
$(m,\ell)$, however it is not completely systematic. In order to fill the gap,
let us use the Frobenius method for solving (\ref{prod1}) \cite{fg07}. With
this aim, let us make the following changes:
\begin{eqnarray}\label{tr2}
&& y=\frac{e_1-\wp(z)}{\bar{e}_2}, \qquad
\Phi(y)=[\wp(z)-e_1]^\ell\Psi.
\end{eqnarray}
Therefore, the following equation for $\Phi$ is obtained:
\begin{eqnarray}
\label{prod2}P_4(y)\frac{d^3\Phi}{dy^3}+P_3(y)\frac{d^2\Phi}{dy^2}+P_2(y)
\frac{d\Phi}{dy}+P_1(y)\Phi = 0,
\end{eqnarray}
where $P_i(y)$ are the following $i$-th degree polynomials in $y$:
\begin{eqnarray}
&& \hskip-0.5cm P_4(y)\!=2y^2(\bar{e}_2y^2-\!3e_1y+\bar{e}_3), \quad
P_3(y)\!=3y[\bar{e}_2(3\!-\!2\ell)y^2\!-\!6e_1(1\!-\!\ell)y+\!\bar{e}_3(1\!-\!2\ell)],
\nonumber \\
&& \hskip-0.5cm P_2(y)=2\{\bar{e}_2(3\ell^2-m^2-6\ell-m+3)y^2
-[\widetilde{E}+e_1(9\ell^2-m^2-9\ell-m+3)]y
           +\ell(2\ell-1)\bar{e}_3\},\nonumber \\
&& \hskip-0.5cm
P_1(y)=\bar{e}_2(2\ell-1)(m+\ell)(m-\ell+1)y+2\ell[\widetilde{E}
+e_1(3\ell^2-m^2-m)]. \nonumber
\end{eqnarray}
Let us propose now
\begin{equation}
\label{fs}\Phi=\sum\limits_{r=0}^\infty a_ry^{r+\rho} ,
\end{equation}
from which the next equations arise
\begin{eqnarray*}
\label{ie} && a_0 f_0(\rho)=0, \quad a_1 f_0(\rho+1)+a_0f_1(\rho) =
0,
\\ \label{re}
&& \hskip-0.5cm
a_{r+2}f_0(\rho+r+2)+a_{r+1}f_1(\rho+r+1)+a_rf_2(\rho+r) = 0, \\
\label{def} && \label{f0} f_0(\rho) =
\bar{e}_3\rho(\rho-1-2\ell)(2\rho-2\ell-1), \quad
\label{f1} f_1(\rho) =  2(\rho-\ell)\{e_1[m(m+1)-3(\rho-\ell)^2]-\widetilde{E}\}, \\
&& \label{f2} f_2(\rho) =
\bar{e}_2(\rho-m-\ell)(\rho+m-\ell+1)(2\rho+1-2\ell).
\end{eqnarray*}
From the three roots $\rho=0, \ \rho=2 \ell + 1, \ \rho = \ell +
1/2$ of the indicial equation $f_0(\rho) = 0$, the series (\ref{fs})
can be made finite just for $\rho=0$, since for $r=2\ell-1$ it turns
out that
\begin{eqnarray}
&& a_{2\ell}f_1(2\ell)+a_{2\ell-1}f_2(2\ell-1)=0 \quad
\Rightarrow \quad \label{a2l} a_r = \frac{(-1)^r
F_r}{\prod\limits^r_{s=1}f_0(s)}a_0 \, ,\quad r=1,2,
                        \dots, 2\ell,
\end{eqnarray}
where the determinant family
\begin{eqnarray}
\label{Fr} F_r \! = \! \left |
   \begin{array}{llllll}
      f_1(r-1)       &      f_2(r-2)        &   \hspace{.7cm} 0    &   \hspace{.7cm} 0   & \cdots & \hspace{.3cm} 0 \\
      f_0(r-1)       &      f_1(r-2)        &       f_2(r-3)       &   \hspace{.7cm} 0   & \cdots & \hspace{.3cm} 0 \\
   \hspace{.7cm} 0   &      f_0(r-2)        &       f_1(r-3)       &      f_2(r-4)       & \cdots & \hspace{.3cm} 0 \\
\hspace{.4cm} \cdots & \hspace{.4cm} \cdots & \hspace{.4cm} \cdots & \hspace{.4cm}\cdots & \cdots &    \cdots       \\
   \hspace{.7cm} 0   &   \hspace{.7cm} 0    &   \hspace{.7cm} 0    &   \hspace{.7cm} 0   & \cdots &    f_1(0)       \\
       \end{array} \right |,
\end{eqnarray}
satisfies the following recurrence relationship:
\begin{eqnarray}
&& F_r = f_1(r-1) F_{r-1} - f_2(r - 2) f_0(r - 1) F_{r-2}, \quad
r\in{\mathbb N}, \quad F_0 \equiv 1, \quad F_{-1} \equiv 0.
\label{recurrencedet}
\end{eqnarray}
In addition, we have to fix $a_{2\ell+1}$ so that the series ends up
after $(N+1)$ terms, i.e.,
\begin{eqnarray*}
\label{are1} & a_Nf_1(N)+a_{N-1}f_2(N-1)=0, \quad a_{N+1}=0, \quad
f_2(N)=0.
\end{eqnarray*}
The previous equations imply that
\begin{eqnarray*}
\label{are2} &  N=m+\ell, \quad
a_{m+\ell}f_1(m+\ell)+a_{m+\ell-1}f_2(m+\ell-1)=0, \quad
a_{m+\ell+1}=0.
\end{eqnarray*}
Two different cases can be identified:

\medskip

\noindent (i) For $m=\ell$ it turns out that formula (\ref{a2l})
provides all non-null $a_r$, $r=0,1,\dots,2\ell$.

\medskip

\noindent (ii) For $m=\ell+\nu,\: \nu=1,2, \ldots$, equation
(\ref{a2l}) supplies just $a_r, \ r=1,\dots,2\ell$, while the
remaining non-null coefficients are given by:
\begin{equation}
\label{fa2l} a_{2\ell+r} = \frac{(-1)^rD_{\nu-r}
\prod\limits_{s=0}^{r-1}f_2(2\ell+s)}{D_{\nu}}a_{2\ell}, \quad
r=1,2,\ldots \nu,
\end{equation}
where $D_r$ is the minor of $F_{2\ell+\nu+1-r}$ in Laplace expansion
of the determinant $F_{2\ell+\nu+1}$.

By coming back now to equation (\ref{tr2}), up to a constant factor
in both cases we have
\begin{eqnarray*}
\label{pp}\Psi(z)=
\frac{\prod\limits_{r=1}^{m+\ell}\left[\wp(z)-\wp(b_r)\right]}{[\wp(z)-e_1]^\ell},
\end{eqnarray*}
$\wp(b_1),\wp(b_2),\ldots, \wp(b_{m+\ell})$ being the zeros of
$\sum\limits_{r=0}^{m+\ell}a_r[(e_1-t)/\bar{e}_2]^r$. By using once
again equation (\ref{usp}), the solutions of the associated Lam\'e
equation are finally:
\begin{eqnarray}
\label{fins} && \hskip-0.5cm \psi^{\pm}(x) \! = \!
\frac{\prod\limits_{r=1}^{m+\ell}\sigma(\frac{x-iK'}{\sqrt{\bar{e}_3}}\pm
b_r)}{\sigma^{\ell}(\frac{x-iK'}{\sqrt{\bar{e}_3}}+\omega_1)
\sigma^{m}(\frac{x-iK'}{\sqrt{\bar{e}_3}})}
\exp\left\{\frac{x}{\sqrt{\bar{e}_3}}\left[\ell \zeta(\omega_1)\mp
\sum\limits_{r=1}^{m+\ell}\zeta(b_r)\right]\right\}.
\end{eqnarray}
Note that the well known solutions of the Lam\'e equation are recovered for
$\ell=0$:
\begin{eqnarray*}
\psi^{\pm}(x)=\frac{\prod\limits_{r=1}^{m}\sigma(\frac{x-iK'}{\sqrt{\bar{e}_3}}\pm
b_r)}{\sigma^{m}(\frac{x-iK'}{\sqrt{\bar{e}_3}})} \exp
\left[\mp\frac{x}{\sqrt{\bar{e}_3}}
\sum\limits_{r=1}^{m}\zeta(b_r)\right].
\end{eqnarray*}

\section{Supersymmetry transformations}

The supersymmetric quantum mechanics, as an approach to generate new
exactly solvable Hamiltonians $\widetilde H$ from an initial
solvable one $H$, is based on the intertwining relation
\begin{equation}
\widetilde H B = B H.
\end{equation}
This means that the eigenfunctions of $\widetilde H$ are constructed
through the non-null action of $B$ onto the eigenfunctions of $H$.
If $B$ is a $k$-th order differential operator it turns out that
\cite{ais93,aicd95}
\begin{equation}
\widetilde V(x)=V(x) - 2[\ln W(u_1,\dots,u_k)]'',
\end{equation}
where $u_i$ are solutions, which can be nonphysical, of the
stationary Schr\"odinger equation associated to $k$ different
factorization energies $\epsilon_i, i=1,\dots,k$, i.e.
\begin{equation}
H u_i = \epsilon_i u_i.
\end{equation}
For periodic potentials, it has been realized that when Bloch-type seed
solutions associated to factorization energies which belong to the energy gaps
are used, then the SUSY partner potentials $\widetilde V(x)$ are again
periodic, isospectral to the initial one \cite{fnn00,fmrs02a,fmrs02b}. On the
other hand, when appropriate linear combinations of the solutions (\ref{fins})
are employed, the SUSY partner potentials $\widetilde V(x)$ of $V(x)$ become
asymptotically periodic, with periodicity defects appearing due to the
creation for $\widetilde H$ of bound states embedded into the gaps
\cite{fmrs02a,fmrs02b}. This suggests a natural ordering for the SUSY
transformations of periodic potentials which will be next followed (we
restrict ourselves to first and second-order techniques).

\subsection{Periodic first-order SUSY partner potentials}

Let us take $u(x) = \psi^{\pm}(x)$ (see (\ref{fins})) with $\epsilon
< E_0$, where $E_0$ is the lowest band edge eigenvalue for the
corresponding associated Lam\'e potential. The periodic first-order
SUSY partner potentials of $V(x)$ thus read
\begin{eqnarray}
\widetilde V_\pm(x) & = & m(m-1)k^2{\rm sn}^2x + \ell (\ell - 1)k^2
\frac{ {\rm cn}^2x}{{\rm dn}^2x} + 2 k^2 \sum\limits_{r=1}^{m+\ell}
{\rm sn}^2(x \pm \sqrt{\bar e_3}b_r). \label{tvp1}
\end{eqnarray}

\subsection{Asymptotically periodic first-order SUSY partner potentials}

Let us choose now $u(x)$ as a general linear combination of the
solutions $\psi^\pm$ ($\epsilon < E_0$),
\begin{eqnarray}\label{linearcombination}
&& \hskip-5pt u(x) = A \psi^+(x) + B \psi^-(x) = A
\psi^+(x)\phi_+(x) = B
\psi^-(x)\phi_-(x), \\
&& \hskip-10pt \phi_\pm(x) \! = \! 1 + \lambda_\pm
\frac{\psi_{\mp}(x)}{\psi_{\pm}(x)} \! = \! 1 + \lambda_\pm
\prod\limits_{r=1}^{m+\ell}\frac{\sigma(\frac{x-iK'}{\sqrt{\bar{e}_3}}\mp
b_r) e^{\pm2 x
\zeta(b_r)/\sqrt{\bar{e}_3}}}{\sigma(\frac{x-iK'}{\sqrt{\bar{e}_3}}\pm b_r)},
\ \lambda_+ = \frac{B}A, \ \lambda_- = \frac{A}B.
\end{eqnarray}
The first-order SUSY partners for the associated Lam\'e potentials
become now asymptotically periodic, with explicit expressions given
by
\begin{equation}
\widetilde V^{np}(x) = \widetilde V_\pm (x) - 2 [\ln \phi_\pm(x)]'',
\label{tvnp1}
\end{equation}
where $\widetilde V_\pm (x)$ is given by (\ref{tvp1}). The spectrum of the
Hamiltonian $\widetilde H^{np}$ contains the allowed energy bands of $H$ but
in addition it has an isolated bound state at $E=\epsilon$.

\subsection{Periodic second-order SUSY partner potentials}

Let us take now two factorization energies $\epsilon_{1,2}$ inside the same
energy gap, and the corresponding Schr\"odinger seed solutions in the way
\begin{eqnarray*}
&& u_{1}(x)=\psi_1^+(x) =
\frac{\prod\limits_{r=1}^{m+\ell}\sigma(\frac{x-iK'}{\sqrt{\bar{e}_3}}+
b_r)}{\sigma^{\ell}(\frac{x-iK'}{\sqrt{\bar{e}_3}}+\omega_1)\sigma^{m}
(\frac{x-iK'}{\sqrt{\bar{e}_3}})}
\exp\left\{{\frac{x}{\sqrt{\bar{e}_3}}\left[\ell \zeta(\omega_1)-
\sum\limits_{r=1}^{m+\ell}\zeta(b_r)\right]}\right\}, \nonumber \\
&& u_{2}(x)= \psi_2^+(x) =
\frac{\prod\limits_{r=1}^{m+\ell}\sigma(\frac{x-iK'}{\sqrt{\bar{e}_3}}+
b'_r)}{\sigma^{\ell}(\frac{x-iK'}{\sqrt{\bar{e}_3}}+\omega_1)\sigma^{m}
(\frac{x-iK'}{\sqrt{\bar{e}_3}})}
\exp\left\{{\frac{x}{\sqrt{\bar{e}_3}}\big[\ell \zeta(\omega_1)-
\sum\limits_{r=1}^{m+\ell}\zeta(b'_r)\big]}\right\}.
\end{eqnarray*}
It turns out that the Wronskian of $u_{1,2}$ is nodeless, which is
conveniently expressed as:
\begin{eqnarray*}
& W(u_1,u_2) =  \psi_1^+(x) \, \psi_2^+(x)\, g(x) \quad g(x) =
\left[\ln\left(\frac{\psi_2^+}{\psi_1^+}\right) \right]'.
\end{eqnarray*}
The second-order SUSY partner potentials of $V(x)$ become periodic:
\begin{eqnarray}
& \hskip-3cm \widetilde V(x) = m(m-3)k^2{\rm sn}^2x + \ell (\ell -
3)k^2 \frac{{\rm cn}^2x}{{\rm dn}^2x} \nonumber \\
& \hskip3cm + 2 k^2\sum\limits_{r=1}^{m+\ell}\left[{\rm sn}^2(x +
\sqrt{\bar e_3}b_r)+ {\rm sn}^2(x + \sqrt{\bar e_3}b'_r)\right] -
2\left(\ln g \right)''. \label{tvp}
\end{eqnarray}

\subsection{Asymptotically periodic second-order SUSY partner potentials}

Let us choose once again $\epsilon_{1,2}$ in the same energy gap but
now $u_{1,2}$ are general linear combinations of $\psi^\pm$,
\begin{eqnarray}\label{lc12}
& u_1(x)=\psi_1^+ + \lambda_1^+ \psi_1^- = \psi_1^+ \phi_1^+, \quad
u_2(x)=\psi_2^+ + \lambda_2^+ \psi_2^- = \psi_2^+ \phi_2^+,
\end{eqnarray}
where
\begin{eqnarray}
&& \phi_1^+ =  1 + \lambda_1^+ \frac{\psi_1^-}{\psi_1^+} = 1 +
\lambda_1^+ \prod\limits_{r=1}^{m+\ell}
\frac{\sigma(\frac{x-iK'}{\sqrt{\bar{e}_3}}-
b_r)e^{2x\zeta(b_r)/\sqrt{\bar{e}_3}}}{\sigma(\frac{x-iK'}{\sqrt{\bar{e}_3}}+
b_r)},
\\
&& \phi_2^+ = 1 + \lambda_2^+ \frac{\psi_2^-}{\psi_2^+} = 1 +
\lambda_2^+ \prod\limits_{r=1}^{m+\ell}
\frac{\sigma(\frac{x-iK'}{\sqrt{\bar{e}_3}}-
b'_r)e^{2x\zeta(b'_r)/\sqrt{\bar{e}_3}}}{\sigma(\frac{x-iK'}{\sqrt{\bar{e}_3}}+
b'_r)}.
\end{eqnarray}
An appropriate choice of $\lambda_{1,2}^+$ leads to a nodeless Wronskian,
which is expressed as:
\begin{eqnarray*}
& W(u_1,u_2) = u_1 u_2 g^{np} = \psi_1^+ \psi_2^+
\phi_1^+ \phi_2^+
g^{np}, \\
& g^{np} = \left[\ln\left(\frac{u_2}{u_1}\right) \right]' =
\left[\ln\left(\frac{\psi_2^+}{\psi_1^+}\right) \right]'+
\left[\ln\left(\frac{\phi_2^+}{\phi_1^+}\right) \right]' = g +
\left[\ln\left(\frac{\phi_2^+}{\phi_1^+}\right) \right]'.
\end{eqnarray*}
The second-order SUSY partner potentials are again asymptotically
periodic:
\begin{eqnarray}
\widetilde V^{np}(x) =  \widetilde V(x) - 2 \left[\ln\left(\phi_1^+
\phi_2^+ \frac{g^{np}}{g}\right) \right]'', \label{tvnp2}
\end{eqnarray}
where $\widetilde V(x)$ is given by (\ref{tvp}).

\section{Example}

In some previous papers it has been studied the associated Lam\'e
potentials and its SUSY partners for $(m,\ell)=(1,1)$,
$(m,\ell)=(2,1)$, and $(m,\ell)=(3,1)$ \cite{ga00,ga02a,fg05,fg07}.
Here we will illustrate our general procedure for the associated
Lam\'e potentials with $(m,\ell)=(3,2)$, i.e.,
\begin{eqnarray}
V(x) = 12k^2{\rm sn}^2x+6k^2\frac{{\rm cn}^2x}{{\rm dn}^2x}.
\end{eqnarray}
Note that there are explicit expressions for the band-edge eigenfunctions and
eigenvalues of $H$ \cite{ga02a}; in particular for the `ground' state it turns
out that:
\begin{eqnarray}
& \psi_0(x)  = {\rm dn}^3x, \quad E_0 = 9 k^2.
\end{eqnarray}

Since $m+\ell=5$, we have to evaluate five constants
$a_1,a_2,a_3,a_4,a_5$, (without losing generality we have taken
$a_0=1$). Let us write down first the basic elements $f_i$,
\begin{eqnarray}\label{s1}
& \hskip-0.5cm f_0(\rho)=\bar e_3\rho(\rho-5)(2\rho-5), \quad
  f_1(\rho)=2(\rho-2)[3e_1\rho(4-\rho)-\widetilde{E}], \\
\label{s2} & f_2(\rho)=\bar e_2(\rho-5)(\rho+2)(2\rho-3), \qquad
  \widetilde{E}=\bar e_3(E-6)+12e_3.
\end{eqnarray}
Note that the four constants $a_1,a_2,a_3,a_4$ are determined from (\ref{a2l})
\begin{equation}\label{s3}
a_1=-\frac{F_1}{f_0(1)}, \ \ a_2=\frac{F_2}{f_0(1)f_0(2)}, \ \
a_3 = - \frac{F_3}{f_0(1)f_0(2)f_0(3)}, \ \ a_4 =
\frac{F_4}{f_0(1)f_0(2)f_0(3)f_0(4)},
\end{equation}
while $a_5$ is computed from (\ref{fa2l})
\begin{equation}\label{s4}
 a_5=-\frac{D_0f_2(4)}{D_1}a_4.
\end{equation}
From (\ref{Fr}-\ref{recurrencedet}) one may calculate quite
straightforwardly the $F_i$'s
\begin{equation}\label{s5}
F_1 = 4 \widetilde E, \quad F_2 = 8(\widetilde E^2 - 9 e_1
\widetilde E - 45 \bar e_2 \bar e_3), \quad F_3 = -288 \bar e_2 \bar
e_3 \widetilde E, \quad F_4 = 5 (72 \bar e_2 \bar e_3)^2 .
                                  \end{equation}
To obtain the $D_i$'s  we need
\begin{equation}
\label{F5} F_6=\left |
   \begin{array}{cccccc}
      f_1(5) & f_2(4) & 0 & 0 & 0 & 0 \\
      f_0(5) & f_1(4) & f_2(3) & 0  & 0 & 0 \\
      0 & f_0(4) & f_1(3) & f_2(2) & 0 & 0 \\
      0 & 0 & f_0(3) & f_1(2) & f_2(1) & 0 \\
      0 & 0 & 0  & f_0(2) & f_1(1) & f_2(0) \\
      0 & 0 & 0  & 0 & f_0(1) & f_1(0)
       \end{array} \right |.
\end{equation}
Then
\begin{eqnarray}\label{s7}
&  D_0=[\mbox{minor of $F_6$ in $F_6$}]=1, \\
\label{s8} & D_1=[\mbox{minor of $F_5$ in $F_6$}]=f_1(5).
\end{eqnarray}
Finally, using the recurrence relation (\ref{recurrencedet}) we
obtain from (\ref{s1}-\ref{s8})
\begin{eqnarray}
&& a_0 = 1, \quad a_1 = -\frac{\widetilde E}{3\bar e_3}, \quad a_2 =
\frac{\widetilde E^2 - 9 e_1 \widetilde E - 45 \bar e_2 \bar
e_3}{9\bar e_3^2}, \nonumber \\ && a_3 = - \frac{2\bar e_2\widetilde
E}{3 \bar e_3^2}, \quad a_4 = \frac{5 \bar e_2^2}{\bar e_3^2}, \quad
a_5 = -\frac{25\bar e_2^3}{\bar e_3^2(\widetilde E+15e_1)}.
\end{eqnarray}
We employ these coefficients to find the roots $c_r, r=1,\dots,5$ of
the fifth-order equation
\begin{equation}
\sum_{r=0}^{5} a_r \left(\frac{e_1-t}{\bar e_2}\right)^r = 0.
\end{equation}
These roots are used then to invert the transcendental equation $\wp(b_r)=
c_r$ to determine the $b_r$'s (with the restriction $\Psi'\vert_{z=b_r}>0$),
which are thus inserted in the explicit expressions for $\psi^{\pm}(x)$.
Finally, the resulting Bloch solutions can be used, either directly or in the
corresponding Wronskian, to derive the periodic SUSY partner potentials
$\widetilde V_\pm(x)$ of (\ref{tvp1}) or (\ref{tvp}). On the other hand,
different linear combinations of kind (\ref{linearcombination}) or
(\ref{lc12}) can be used to derive the potentials $\widetilde V^{np}(x)$ of
(\ref{tvnp1}) or (\ref{tvnp2}), which have periodicity defects. The final
results of these procedures are illustrated in figures below, in which we show
in gray the original associated Lam\'e potential for $m=3$, $\ell = 2$,
$k^2=0.9$. In Figure 1a we plot in black one of its periodic first-order SUSY
partners generated through a Bloch solution with $\epsilon = 8 < E_0 = 8.1$,
while in Figure 1b it is illustrated one of its asymptotically periodic partners for the
same $\epsilon$. On the other hand, in Figure 2 we have drawn similar graphs
(black curves) for the corresponding second-order SUSY partner potentials,
periodic and asymptotically periodic. For the periodic case (Figure 2a) we
have used two Bloch solutions $u_1(x) = \psi_1^+(x)$, $u_2(x) = \psi_2^+(x)$
associated to $\epsilon_1 = 10$ and $\epsilon_2 = 10.1$, which fall in the
first finite energy gap $(8.1031,11.7154)$. For the asymptotically periodic
case (Figure 2b) we have used the same pair of factorization energies, with
linear combinations $u_1(x) = \psi_1^+(x) + \psi_1^-(x)$ and $u_2(x) =
\psi_2^+(x)-1.5 \ \psi_2^-(x)$. Note that in both asymptotically periodic
cases, of first and second order, the periodicity defects are clearly
detected.

\begin{figure}[ht]
\epsfig{file=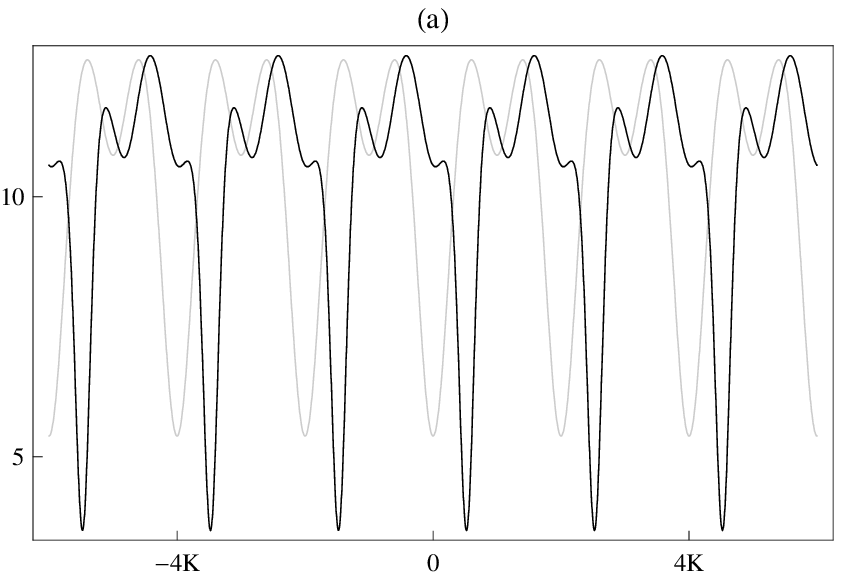,width=7cm}\hskip1cm
\epsfig{file=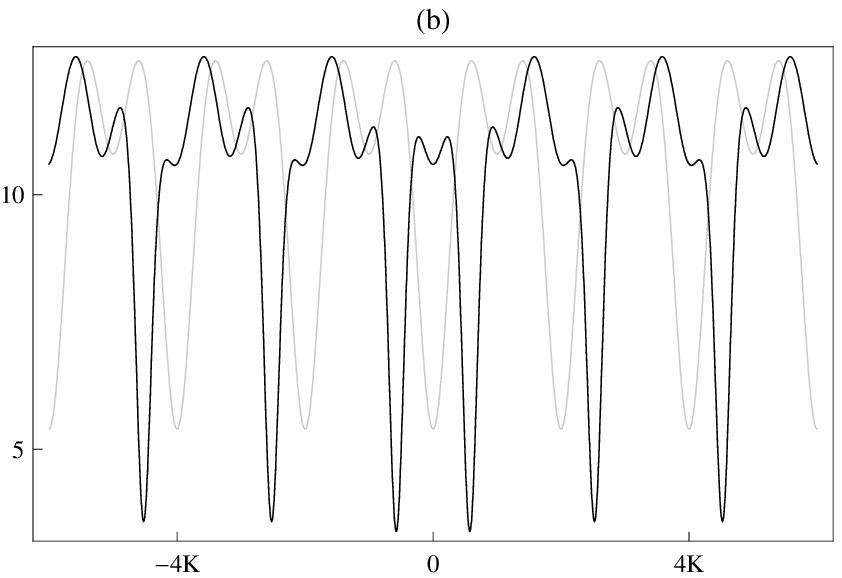,width=7cm} \caption{\small First-order SUSY
partners (black curves) for the associated Lam\'e potential (gray
curves) with $m=3, \ \ell=2$, $k^2=0.9$. (a) Periodic case generated
through the Bloch solution $\psi^+(x)$ with $\epsilon = 8$. (b)
Asymptotically periodic case in which the linear combination
$\psi^+(x) + \psi^-(x)$ for the same $\epsilon$ is used.}
\end{figure}

\section{Conclusions}

We have shown that the associated Lam\'e potentials for integers
values of the parameter pair $(m,\ell)$ are exactly solvable. When
using as seeds Bloch-type solutions inside the gaps new exactly
solvable periodic potentials are generated. On the other hand, for
seeds chosen as general linear combinations of Bloch-type solutions,
the SUSY partner potentials become asymptotically periodic, the
corresponding spectra having bound states embedded into the gaps.
These potentials are interesting since the new levels could work as
intermediate transition energies for the electrons to jump between
the energy bands.

\begin{figure}[ht]
\epsfig{file=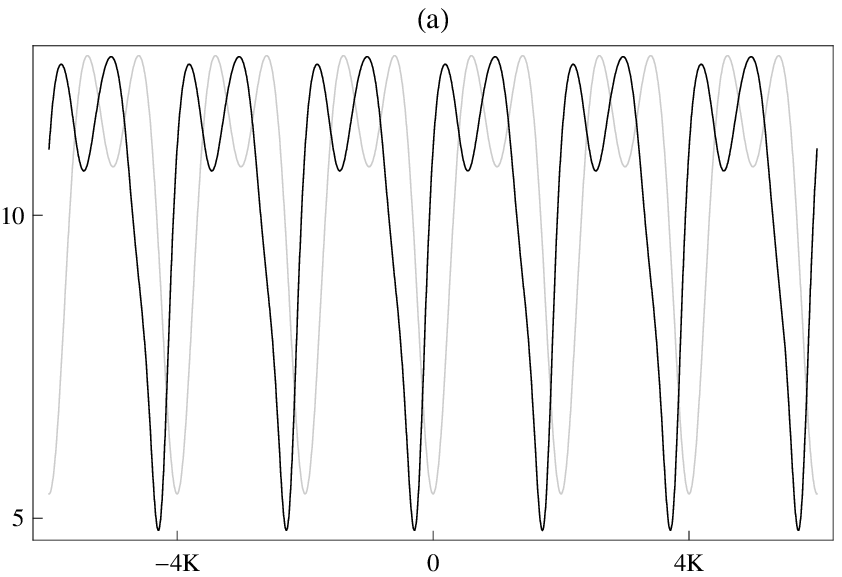, width=7cm} \hskip1cm \epsfig{file=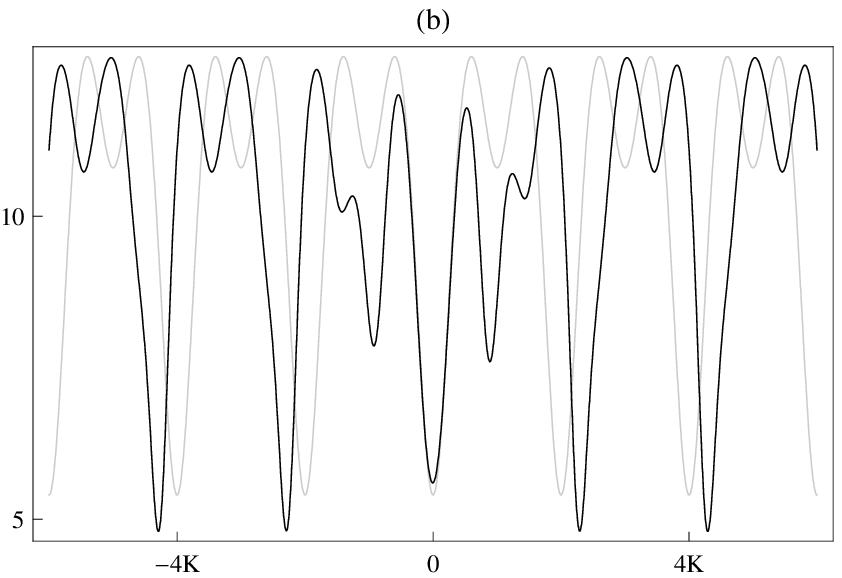,
width=7cm} \caption{\small Second-order SUSY partners (black curves)
for the associated Lam\'e potential (gray curves) with $m=3, \
\ell=2$, $k^2=0.9$. (a) Periodic case generated through two Bloch
solutions $u_1(x) = \psi_1^+(x)$, $u_2(x) = \psi_2^+(x)$ for
$\epsilon_1 = 10$, $\epsilon_2 = 10.1$. (b) Asymptotically periodic
case in which the linear combinations $u_1(x) = \psi_1^+(x) +
\psi_1^-(x)$, $u_2(x) = \psi_2^+(x)-1.5 \ \psi_2^-(x)$ for the same
$\epsilon_{1,2}$ are used.}
\end{figure}

We would like to end up this paper by making a historical precision concerning
SUSY techniques applied to periodic potentials (see also the discussion in
\cite{cjp08}). It is a fact that the recent interest on the subject
\cite{df98,ks99,fnn00,fmrs02a,fmrs02b,sgnn03,fg05,fg07,gin06,imv08,cjnp08} was
catalyzed by Dunne and Feinberg discovery of the self-isospectrality for the
Lam\'e potentials with $m=1$, induced by the first-order SUSY transformation
which employs as seed the ground-state eigenfunction. Self-isospectrality
means, in particular, that the SUSY partner potential becomes just a displaced
version of the initial one, by half the period in \cite{df98}. Soon it was
realized that the self-isospectrality in which the new potential becomes the
initial one displaced by any real number arises as well for the Lam\'e
potential with $m=1$, the seeds employed being Bloch-type solutions associated
to factorization energies in the gaps \cite{fmrs02a,fmrs02b}. More general
SUSY transformations, either of higher order or involving general solutions of
the Schr\"odinger equation for a given factorization energy, have been also
introduced \cite{fmrs02a,fmrs02b,sgnn03}. However, it is worth to note that
there are several interesting works, previous to Dunne and Feinberg paper, in
which the SUSY techniques were applied to periodic potentials (see e.g.
\cite{tr89,bm85}). In particular, it is remarkable the work of Braden and
Macfarlane in which the self-isospectrality for the Lam\'e potential with
$m=1$ was discovered for the first time \cite{bm85}. This is a typical story
of a discovery followed by a later rediscovery, which often arises in science.
Our opinion is that both works are valuable, complementary to each other, and
hence they are worth to be studied in detail.

\section*{Acknowledgments} The authors acknowledge the support of
Conacyt, project No. 49253-F.

\end{document}